\documentclass[aps,pre,twocolumn,superscriptaddress,groupedaddress]{revtex4} 
\usepackage{graphicx}
\usepackage{dcolumn}
\usepackage{bm}
\usepackage[hidelinks]{hyperref}
\usepackage[version=3]{mhchem} 
\usepackage[T1]{fontenc}       
\usepackage{siunitx}
\usepackage{amssymb}
\usepackage{amsmath}

\begin{document}

\author{Pierre Lidon}
\affiliation{Univ Lyon, Ens de Lyon, Univ Claude Bernard, CNRS, Laboratoire de physique, F-69342 Lyon, France}
\affiliation{CNRS, Solvay, LOF, UMR 5258, Univ. Bordeaux, 178 avenue du Dr. Schweitzer, F-33600 Pessac, France}
\email{pierre.lidon@u-bordeaux.fr}

\author{Louis Villa}
\author{S{\'e}bastien Manneville}
\affiliation{Univ Lyon, Ens de Lyon, Univ Claude Bernard, CNRS, Laboratoire de physique, F-69342 Lyon, France}

\title{A mesoscale study of creep in a microgel using the acoustic radiation force}

\begin{abstract}
We study the motion of a sphere of diameter 330~\SI{}{\micro\meter} embedded in a Carbopol microgel under the effect of the acoustic radiation pressure exerted by a focused ultrasonic field. The sphere motion within the microgel is tracked using videomicroscopy and compared to conventional creep and recovery measurements performed with a rheometer. We find that under moderate ultrasonic intensities, the sphere creeps as a power law of time with an exponent $\alpha\simeq 0.2$ that is significantly smaller than the one inferred from global creep measurements below the yield stress of the microgel ($\alpha\simeq 0.4$). Moreover, the sphere relaxation motion after creep and the global recovery are respectively consistent with these two different exponents. By allowing a rheological characterization at the scale of the sphere with forces of the order of micronewtons, the present experiments pave the way for acoustic ``mesorheology'' which probes volumes and forces intermediate between standard macroscopic rheology and classical microrheology. They also open new questions about the effects of the geometry of the deformation field and of the sphere size and surface properties on the creep behaviour of soft materials.
\end{abstract}

\maketitle



\section{Introduction}
\label{sec:introduction}

Soft matter encompasses a large variety of physico-chemical systems, displaying a wide range of mechanical behaviours, intermediate between solid and liquid. These complex fluids are ubiquitous in natural and industrial contexts and it is thus of crucial importance to understand and control their flow properties.

Among the large variety of ways to induce flow, the motion of spherical objects, such as droplets or solid particles, under a constant force in complex fluids has been extensively studied due to its conceptual simplicity and relevance to many practical situations. In viscoelastic fluids, e.g. polymer and micellar solutions, the settling of spheres has been described in depth through many experiments \cite{bisgaard_1983,bush_1993,bush_1994,arigo_1995,solomon_1996,arigo_1997,handzy_2004,verneuil_2007,mohammadigoushki_2016} and is now used as a benchmark to test numerical models \cite{bush_1993,bush_1994,arigo_1995,feng_1996,arigo_1997}. In particular, the flow patterns observed for large deformations in the nonlinear regime is associated with complex wake instabilities\cite{harlen_1990,arigo_1998,fabris_1999,jayaraman_2003,chen_2004} due to shear-thinning properties.

Yield stress fluids are a class of soft materials that behave as solids for stresses below a critical value $\sigma_\text{y}$, the yield stress, but flow like liquids for stresses larger than $\sigma_\text{y}$\cite{larson_1999,barnes_1999,coussot_2005,mewis_2012} (see also Ref.~\cite{bonn_2017} for a recent review). In that sense, yield stress fluids display a diverging viscosity close to the yield stress and modeling the motion of a sphere within a yield stress fluid appears as particularly challenging\cite{coussot_2014,balmforth_2014,chaparian_2017a}. Previous experimental studies have mainly focused on fluid motion induced by dragging large spheres through yield stress fluids either at constant speed\cite{jossic_2001,tokpavi_2009,boujlel_2012a,boujlel_2012b,chevalier_2013b,debruyn_2004,chafe_2005} or at constant force\cite{atapattu_1995,hariharaputhiran_1998,tabuteau_2007,putz_2008,gueslin_2009,sikorski_2009,holenberg_2012}. In this case, complex flow patterns and memory effects were observed. Numerical simulations provide reasonable agreement with experimental results for large stresses \cite{beris_1985,blackery_1997,beaulne_1997,liu_2002,liu_2003,tokpavi_2008,putz_2010,fraggedakis_2016} but generally fail at describing the behaviour close to the yield stress, as they rely on non-physical regularizations of the constitutive equation for small stresses \cite{papanastasiou_1987}. 

No local study is yet available for spheres subjected to some stress below $\sigma_\text{y}$. Still, yield stress fluids are known to exhibit interesting slow motion below $\sigma_\text{y}$, namely creep characterized by ever-decreasing yet finite deformation rates \cite{plazek_1960,paredes_2013,siebenburger_2012,grenard_2014,leocmach_2014,lidon_2017a}. The consequences of such creeping deformations on the motion of settling objects has not been studied. They however question the stability of suspensions of particles and droplets in yield stress fluids, which has been suggested as a path to create soft solids with tunable rheology\cite{kogan_2013,ducloue_2014,ducloue_2015,style_2015a,style_2015b,koblitz_2018}. Moreover, previous experiments on the motion of objects in complex fluids mainly studied the displacement of macroscopic spheres under gravitational forces. Yet, buoyancy is not easy to tune for the study of smaller particles, relevant in the context of suspensions and bubbly gels. Micromanipulation methods like optical\cite{starrs_2003,gutsche_2008,sriram_2009,gomezsolano_2014,gomezsolano_2015} and magnetic\cite{ziemann_1994,amblard_1996,wilhelm_2002a,wilhelm_2008,choi_2011,rich_2011,cribb_2013} tweezers allowed recent progress on the understanding of the motion of small objects in complex fluids. However, these techniques are limited to forces up to about a few nanonewtons and to particles of size between $\SI{100}{\nano\meter}$ and $\SI{10}{\micro\meter}$. Thus they provide access to the behaviour close to the yield stress only for very soft gels.

In this paper, we propose to use the acoustic radiation force to push on a small sphere embedded in a Carbopol microgel. It has been recognized for decades \cite{wood_1927,hertz_1939,goldman_1949} that when submitted to intense ultrasonic waves, any interface between two media with different acoustic impedances undergoes a steady force, known as the acoustic radiation pressure. In a previous paper \cite{lidon_2017b}, we showed that this nonlinear effect allows one to exert a remote force above a few micronewtons on a polystyrene sphere of diameter $\sim\SI{300}{\micro\meter}$ embedded in a Carbopol microgel. By exploiting the elastic regime of deformation at short times and relatively low ultrasonic intensities, we were able to calibrate the intensity of the acoustic force. Here, we analyze in more details the long-time plastic deformation following this initial elastic regime as well as the relaxation of the sphere after removal of the force. We show that the creep motion of the sphere under the effect of the acoustic radiation force is well described by a power law. The relaxation motion of the sphere once the force is removed is consistent with the same power-law exponent.
    
We further study creep and recovery in the same microgel by use of conventional rheology and show that the deformation is well described by power laws as observed with our acoustical technique yet with a significantly larger exponent. Important discrepancies make it hard to directly and quantitatively compare both experiments. First and foremost, the geometry of the deformation field around the sphere in the acoustical experiments is not controlled and presumably more complex than the simple shear investigated with a rheometer. Second, the nature of the moving surfaces and the probed time and length scales are different. Still, beyond the comparison between acoustic measurements and classical rotating rheometry, the present work opens a promising path towards a new method of acoustic ``mesorheology'' that should bring about new insights on the local dynamics of Carbopol microgels below and close to the yield stress at a mesoscale of a few hundreds of micrometers.


\section{Materials and methods}
\label{sec:methods}

\subsection{Carbopol microgels}
\label{subsec:carbopol}

Carbopol is a mixture of polymeric particles of cross-linked acrylic acid. When dispersed in water and neutralized, polymer particles swell and form a soft jammed assembly of elastic coils, with a typical size of a few micrometers \cite{baudonnet_2004,piau_2007,lee_2011,gutowski_2012,geraud_2013,geraud_2017}, that is referred to as a microgel. Carbopol microgels are frequently described as model yield stress fluids, as their rheological behaviour is satisfactorily described by the Herschel-Bulkley constitutive equation above the yield stress \cite{bonn_2017} and as they do not display significant aging, contrary to many other systems \cite{piau_2007,gutowski_2012}. However, for stresses slightly above the yield stress, some Carbopol microgels may display complex fluidization dynamics, associated with delayed yielding \cite{divoux_2011b}, stress overshoots \cite{divoux_2011a} and transient shear banding \cite{divoux_2010}. They also quite generally display creep deformation and residual stresses below the yield stress \cite{lidon_2017a}. The rheology of Carbopol microgels close to the yield stress is thus challenging and calls for local investigations. Here, we study a microgel made of Carbopol ETD~2050 at a mass fraction of $1\%$, prepared using the procedure described in Refs.~\cite{geraud_2013,geraud_2017,lidon_2017a}. 

\subsection{Rheological setup}
\label{subsec:rheol}

\begin{figure}[t]
\centering
  \includegraphics[height=5cm]{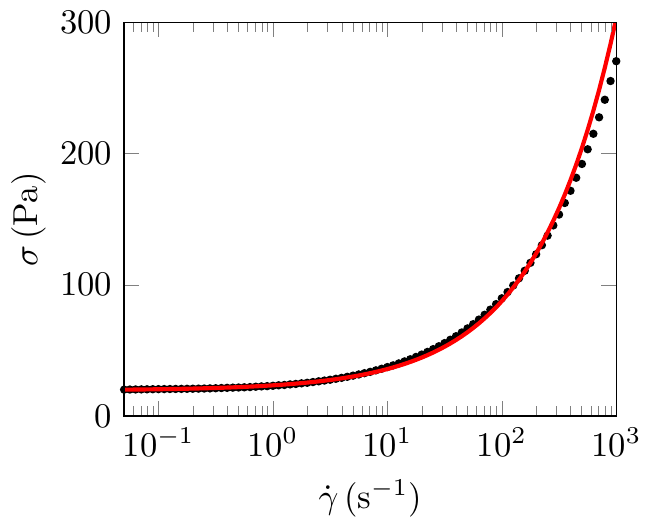}
  \caption{Flow curve, shear stress $\sigma$ vs. shear rate $\dot{\gamma}$, of a 1\%~wt. Carbopol microgel, measured by decreasing the shear rate from $\SI{1000}{\per\second}$ in logarithmically-spaces steps of $\SI{5}{\second}$ each. The red line is a fit by the Herschel-Bulkley law [Eq.~\eqref{eq:herschel_bulkley}] with $\sigma_\text{y}=\SI{19.5}{\pascal}$, $n=0.62$ and $k=3.92 \, \text{SI}$.}
  \label{fig:Carbopol_rheology}
\end{figure}

The {\it macroscopic} rheology of our microgel is measured with a stress-controlled rheometer (Anton Paar MCR301) equipped with a roughened cone-and-plate geometry (sandblasted cone of diameter $\SI{40}{\milli\meter}$ and angle $\SI{2.0}{\degree}$) and a smooth bottom plate including a Peltier element that imposes the temperature to $\SI{25}{\degree}$C. The geometry is covered by a homemade lid to minimize evaporation, allowing for measurements on the same sample for about 12~h. Figure~\ref{fig:Carbopol_rheology} displays the flow curve of the microgel under study. As expected for a simple yield stress fluid, it is accurately described by the Herschel-Bulkley law
\begin{equation}
\sigma = \sigma_\text{y} + k \dot{\gamma}^n\,,
\label{eq:herschel_bulkley}
\end{equation}
\noindent with a yield stress $\sigma_\text{y}=\SI{19.5}{\pascal}$, an exponent $n=0.62$ and a consistency index $k=3.92 \, \text{SI}$. In the following, we shall focus on creep tests performed below the yield stress $\sigma_\text{y}$ and followed by recovery tests. To provide a full rheological characterization of our sample, Fig.~\ref{fig:Carbopol_viscoelasticity} in Appendix~A shows the viscoelastic moduli of the microgel measured both in the linear and nonlinear regimes.

\subsection{Acoustical setup}
\label{subsec:setup}

\begin{figure}[t]
\centering
  \includegraphics[width=\columnwidth]{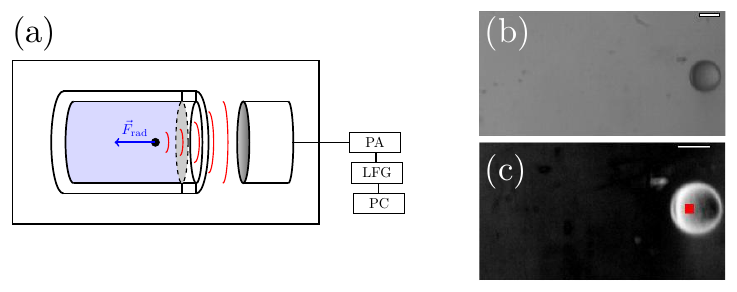}
  \caption{Schematics of the acoustical experimental setup and image analysis. (a)~The piezoelectric transducer is powered by a power-amplifier (PA) driven by a function generator (LFG) controlled by a computer (PC). A polystyrene sphere, embedded in the microgel under study, is placed at the focus of the transducer and its motion under the acoustic radiation force $\vec{F}_\text{rad}$ is recorded by a fast CCD camera. (b)~Image of a polystyrene sphere embedded in a 1\%~wt. Carbopol microgel. Scale bar is $\SI{200}{\micro\meter}$. (c) Image analysis: the red square is the detected position. Uncertainty on the position is about $\SI{5}{\micro\meter}$. Scale bar is $\SI{200}{\micro\meter}$.}
  \label{fig:setup}
\end{figure}

\paragraph*{Experimental cell.} The acoustical setup used to probe the microgel at a {\it mesoscale} is sketched in Fig.~\ref{fig:setup}(a) and is identical to the one described in Ref.\cite{lidon_2017b}. A transparent cylindrical cell of inner diameter $\SI{2}{\centi\meter}$ and length $\SI{5}{\centi\meter}$ is filled with a Carbopol microgel and a single polystyrene sphere of radius $a=163 \pm 3 \, \SI{}{\micro\meter}$ is introduced within the microgel. Air patches generated in this process are removed using a syringe. The cell is then closed with a thin plastic wrap, taking great care not to trap any air bubble that would disturb the ultrasonic field. The density mismatch between the Carbopol microgel ($\rho_\text{g} \simeq \SI{1.03e3}{\kilogram\per\meter\cubed}$) and the polystyrene sphere ($\rho_\text{p} \simeq \SI{1.05e3}{\kilogram\per\meter\cubed}$) is small enough to ensure that the buoyant stress, $\sigma_\text{b}\simeq ag (\rho_\text{p}-\rho_\text{g}) \simeq\SI{0.03}{\pascal}$ with $g$ the gravity, is much smaller than the yield stress, $\sigma_\text{y}\simeq\SI{20}{\pascal}$, which prevents any settling of the sphere during the experiment.

\paragraph*{High-intensity focused ultrasound.} Ultrasound is generated by a hemispherical piezoelectric transducer (Imasonics, diameter $\SI{38}{\milli\meter}$, central frequency $f=\SI{2.25}{\mega\hertz}$), powered by a broadband power amplifier (Kalmus 150C) driven by a function generator (Agilent 33522A). The transducer generates an acoustic field that is axisymmetric around the axis of propagation and focused at $\SI{38}{\milli\meter}$ from the transducer. The focal spot, over which the acoustic pressure can be considered as homogeneous under free propagation,\cite{lidon_2017b} has a diameter of $\SI{1.0}{\milli\meter}$ and an extension $\ell = \SI{5}{\milli\meter}$ along the acoustic propagation axis. The pressure amplitude at focus $P_0$ is controlled to values up to $\sim\SI{7}{\mega\pascal}$. The cell and the transducer are immersed in a large water tank at room temperature 20--$\SI{25}{\celsius}$ to ensure ultrasound propagation. With the sound speed in water $c=\SI{1500}{\meter\per\second}$ at room temperature, the acoustic wavelength $\lambda = c / f = \SI{670}{\micro\meter}$ is about twice as large as the sphere diameter of $\SI{330}{\micro\meter}$.

\paragraph*{Sphere displacement.} Acoustic waves are focused on the polystyrene sphere initially located at the focal spot and its motion under the effect of the acoustic radiation pressure is recorded at $300\, \text{fps}$ with a fast CCD camera (Baumer HXC20) mounted on a macroscope (Nikon SMZ745T). The cell is illuminated by a LED panel ensuring a homogeneous bright background. The position $r(t)$ of the sphere is obtained by standard image analysis as displayed on Fig.~\ref{fig:setup}(b,c): tracking a spot that remains fixed on the sphere surface allows us to locate the sphere with a precision of about $\SI{5}{\micro\meter}$.

\paragraph*{Force calibration.} For low acoustic intensities, we may neglect long-time creep and model the microgel by a purely elastic material of shear modulus $G_0$ so that the displacement $\delta r$ of a sphere of radius $a$ under the acoustic radiation force $F_\text{rad}$ is given by\cite{ilinskii_2005,urban_2011}
\begin{equation}
\delta r = \frac{F_\text{rad}}{6\pi a G_0}\,.
\label{eq:force_displacement}
\end{equation}
In our previous paper~\cite{lidon_2017b}, we demonstrated experimentally that the displacement of the sphere at low acoustic intensities is proportional to the square of the acoustic pressure at focus $P_0^2$ in the absence of the sphere. Together with Eq.~\eqref{eq:force_displacement}, this provides the calibration of $F_\text{rad}$ in terms of the experimental control parameter $P_0$, which is tuned with the voltage delivered by the amplifier. Here, using the value $G_0=G_\text{c}=\SI{51.3}{\pascal}$ inferred from creep measurements (see Sect.~\ref{subsec:results_creep} below), we obtain $F_\text{rad} = \beta P_0^2$ with $\beta = \SI{2.8}{\micro\newton\per\mega\pascal\squared}$. The reader is referred to Ref.~\cite{lidon_2017b} for more details on the force calibration.

\paragraph*{Homogeneity of the acoustic radiation force field.} As previously reported~\cite{lidon_2017b}, successive reflections of the acoustic wave between the sphere and the transducer lead to the formation of an acoustic cavity, hence creating spatial oscillations of the acoustic pressure with a wavelength $\lambda/2=\SI{335}{\micro\meter}$. Thus, the acoustic radiation force field can no longer be considered as strictly homogeneous for displacements larger than a few tens of micrometers. As further discussed below in Sects.~\ref{subsec:preliminary} and \ref{subsec:disc_elastic_param}, this acoustic Fabry-P{\'e}rot effect somewhat complicates the interpretation of $r(t)$ at large acoustic intensities.

\section{Experimental results}
\label{sec:results}

In this section, we compare results from standard rheological measurements and from our local measurements of the motion of a sphere driven by the acoustic radiation pressure. We first describe the experimental protocols (Sect.~\ref{subsec:protocol}) as well as preliminary observations performed with our acoustical setup (Sect.~\ref{subsec:preliminary}). Creep and recovery measurements are then detailed in Sects.~\ref{subsec:results_creep} and \ref{subsec:results_recovery} respectively. The results are further discussed in Sect.~\ref{sec:discussion}.

\subsection{Protocols for rheological and acoustical measurements}
\label{subsec:protocol}

In conventional rheology, the microgel is first presheared at a shear rate $\dot{\gamma} = \SI{100}{\per\second}$ during $\SI{60}{\second}$ prior to each creep and recovery test. It is then left to rest under zero deformation for $\SI{300}{\second}$. During these first two steps, strain is controlled through the feedback loop of our stress-imposed rheometer. Finally, a constant shear stress $\sigma$ is applied during $\SI{300}{\second}$ (creep phase) and set back to zero (recovery phase). The strain $\gamma$ is recorded by the rheometer as function of time $t$ during both phases. Such macroscopic rheological measurements provide us with a basis to analyze and interpret the displacement of a sphere under the acoustic radiation force.

In our acoustical setup, it is obviously more difficult to ensure a reproducible initial state than in a rheometer. Still, the stirring generated when filling the cell and when removing air bubbles should be sufficient to erase memory of previous deformations in the microgel. We successively switch the ultrasonic excitation on and off so that the sphere displacement $r(t)$ is recorded during two phases that mimic the conventional creep and recovery experiment, yet on significantly shorter timescales due to the limitations of our high-power transducer. Indeed, ultrasound is first turned on during $\SI{3}{\second}$ and the sphere is pushed by the acoustic radiation force in the direction of propagation of the ultrasonic beam (``creep'' phase). Then, the ultrasonic excitation is switched off for about one minute (``recovery'' phase). During this phase, the sphere is pulled backwards due to the relaxation of the microgel deformation. The durations of the creep and recovery phases are chosen such that the transducer does not heat up significantly during the creep phase and such that the sphere reaches a stationary position during the recovery phase.

\subsection{Preliminary acoustical measurements}
\label{subsec:preliminary}

\begin{figure}[b]
\centering
  \includegraphics[width=\columnwidth]{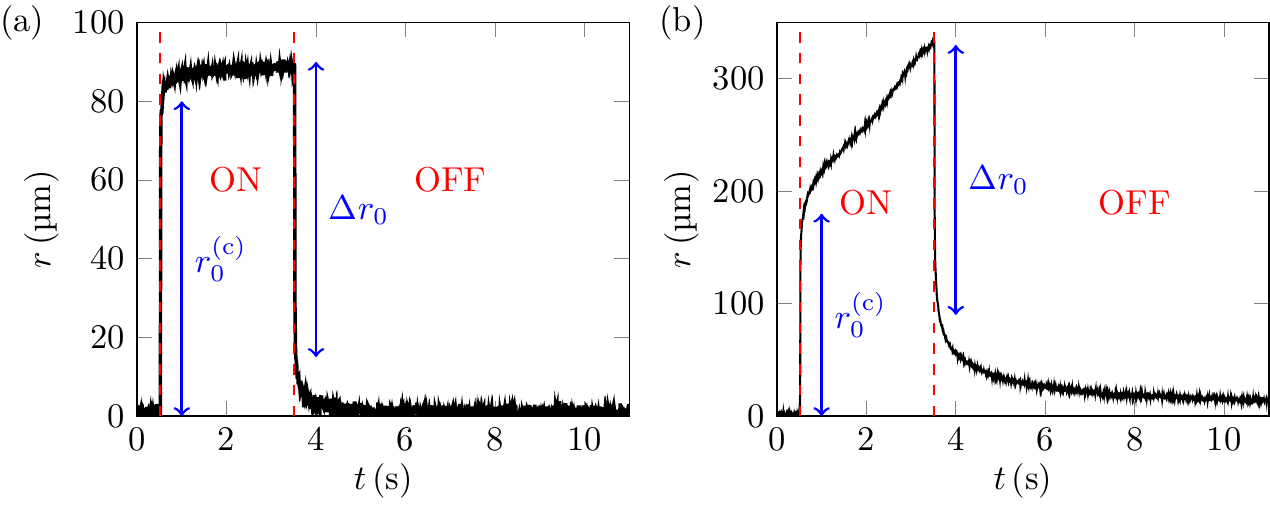}
  \caption{Displacement $r$ of the sphere as a function of time $t$ under the effect of the acoustic radiation force. Acoustic excitation is turned on during $\SI{3}{\second}$. The polystyrene sphere is embedded in a 1\%~wt. Carbopol microgel. (a) $F_\text{rad}=\SI{13.2}{\micro\newton}$: the excitation is below the yield stress and the sphere displays creep motion. (b) $F_\text{rad}=\SI{32.1}{\micro\newton}$: the acoustic force overcomes the yield value $F_\text{y}$ and the sphere fluidizes the surrounding microgel.}
  \label{fig:example_US}
\end{figure}

\begin{figure*}[t]
\centering
  \includegraphics[height=10cm]{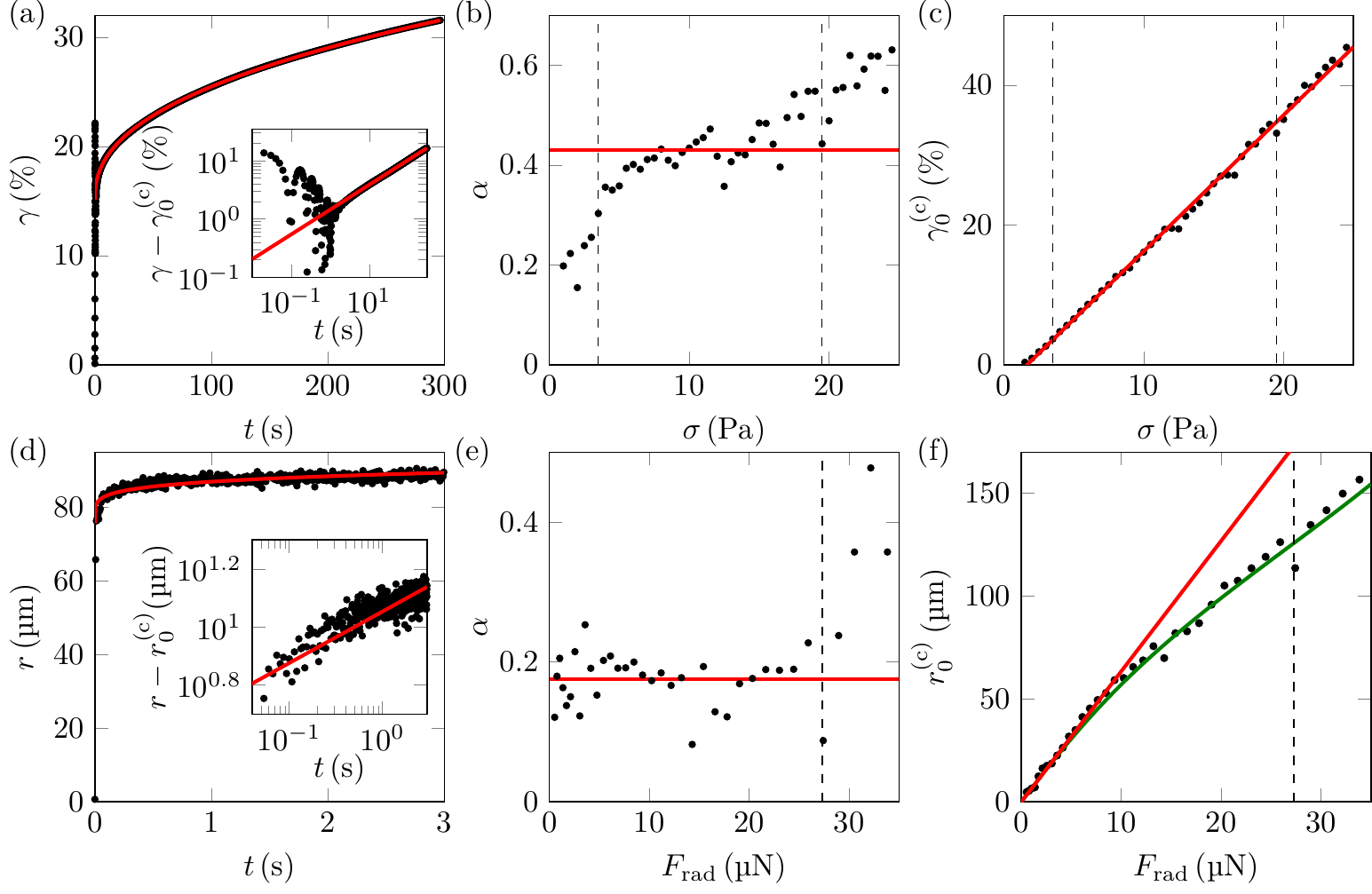}
  \caption{(a) Creep of a 1\%~wt. Carbopol microgel as measured with a rheometer in a cone-and-plate geometry under a constant stress $\sigma = \SI{9.5}{\pascal} \simeq 0.5\sigma_\text{y}$. The red curve is a fit of the strain response $\gamma(t)$ by the Andrade law [Eq.~\eqref{eq:andrade}] with $\gamma^\text{(c)}_0=15.1\%$, $\alpha=0.43$ and $\tau=\SI{0.41}{\second}$. The inset shows $\gamma-\gamma^\text{(c)}_0$ as a function of time $t$ in logarithmic scales. (b)~Evolution of the Andrade exponent $\alpha$ with the applied stress $\sigma$. The red line indicates the average value $\langle \alpha \rangle = 0.43$. (c)~Instantaneous elastic deformation $\gamma^\text{(c)}_0$ as a function of $\sigma$. The red line is a linear fit, $\gamma^\text{(c)}_0 =  (\sigma - \sigma_{0,\text{c}})/ G_\text{c}$, with $G_\text{c} = \SI{51.3}{\pascal}$ and $\sigma_{0,\text{c}}=\SI{1.65}{\pascal}$. The vertical dashed lines in (b) and (c) indicate the range of stresses over which the Andrade scaling correctly accounts for the $\gamma(t)$ data, from $2\sigma_{0,\text{c}}\simeq\SI{3.5}{\pascal}$ to the yield stress $\sigma_\text{y}=19.5$~Pa. (d)~Displacement of a polystyrene sphere embedded in a 1\%~wt. Carbopol microgel under an acoustic radiation force $F_\text{rad}=\SI{13.2}{\micro\newton}\simeq 0.5 F_\text{y}$. The red curve is a fit of $r(t)$ to the power-law behaviour of Eq.~\eqref{eq:andrade_US} with parameters $r_0^\text{(c)} = \SI{75.7}{\micro\meter}$, $\alpha = 0.18$ and $\tau=\SI{4.5e4}{\second}$. The inset shows $r-r^\text{(c)}_0$ as a function of time $t$ in logarithmic scales. (e)~Evolution of the Andrade exponent $\alpha$ with the acoustic radiation force $F_\text{rad}$. The red line indicates the average value $\langle \alpha \rangle = 0.18$ for $F_\text{rad}<F_\text{y}=\SI{27.3}{\micro\newton}$. (f)~Instantaneous elastic displacement $r_0^\text{(c)}$ of the sphere as a function of $F_\text{rad}$. The red line is a linear fit, $r_0^\text{(c)} = \kappa_\text{c} F_\text{rad}$ for $F_\text{rad}<\SI{11}{\micro\newton}$, with $\kappa_\text{c} = 6.3\pm\SI{0.3}{\meter\per\newton}$. The green curve is the theoretical prediction accounting for the local variations of the pressure field due to an acoustic cavity effect between the sphere and the transducer (see text in Sect.~\ref{subsec:disc_elastic_param}). The vertical dashed line in (e) and (f) indicates $F_\text{y}=\SI{27.3}{\micro\newton}$.}
  \label{fig:creep}
\end{figure*}

Figure~\ref{fig:example_US} shows two typical measurements of the sphere displacement as a function of time. Figure~\ref{fig:example_US}(a) corresponds to a moderate value of the acoustic radiation force. In this case, when ultrasound is switched on, the sphere first moves instantaneously forward by a distance $r^\text{(c)}_0$ and then reaches an almost stationary position, with small and slow creep motion. When the acoustic force is removed, the sphere instantaneously recoils by a distance $\Delta r_0$ and then relaxes towards its initial position.

For forces larger than a threshold value $F_\text{y}=\SI{27.3}{\micro\newton}$, the velocity of the sphere does not vanish after the initial elastic deformation as indicated by the positive slope of $r(t)$ in Fig.~\ref{fig:example_US}(b). After the end of ultrasonic excitation, the sphere does not come back to its initial position either. This demonstrates that irreversible deformations occur within the microgel due to flow above the yield stress. When such irreversible displacements come into play, they may accumulate from one measurement to the other. This rapidly leads to displacements comparable to $\lambda/2$, making it difficult to conduct reproducible experiments due to the above-mentioned acoustic cavity effect. Therefore, we shall mostly focus on the regime of moderate acoustic forces, up to $F_\text{y}$, where deformations remain essentially reversible.

\subsection{Creep experiments}
\label{subsec:results_creep}

Typical results of creep experiments are displayed in Fig.~\ref{fig:creep}(a) for rheological measurements and in Fig. \ref{fig:creep}(d) for acoustical measurements. We now proceed with a thorough investigation of both creep behaviours as a function of their respective control parameter, namely the shear stress $\sigma$ and the acoustic radiation force $F_\text{rad}$. 

\paragraph*{Rheological data.} The creep deformation of ETD~2050 Carbopol microgels has been studied in much details in a previous paper\cite{lidon_2017a}. Data that we report here on the Carbopol microgel used for acoustical experiments are in full agreement with our previous conclusions. A typical creep test under an imposed stress $\sigma\simeq 0.5\sigma_\text{y}$ is displayed in Fig.~\ref{fig:creep}(a). Once the short-time inertio-elastic oscillations are completely damped~\cite{baravian_2007,caton_2008,benmouffok_2010}, the growth of the strain $\gamma$ with time $t$ is well fitted by a power law, also referred to as Andrade creep\cite{andrade_1910},
\begin{equation}
\gamma (t) = \gamma^\text{(c)}_0\left[ 1 + \left( \frac{t}{\tau} \right)^\alpha \right]\,,
\label{eq:andrade}
\end{equation}
where the exponent $\alpha \simeq 0.4$ weakly increases with the applied stress $\sigma$ as already observed in Ref.\cite{lidon_2017a} [see Fig.~\ref{fig:creep}(b)]. Moreover, as shown in Fig.~\ref{fig:creep}(c), the instantaneous deformation $\gamma^\text{(c)}_0$ evolves linearly with the stress following $\gamma^\text{(c)}_0 = (\sigma - \sigma_{0,\text{c}})/ G_\text{c}$. The slope $G_\text{c} = \SI{51.3}{\pascal}$ defines an effective elastic modulus that we use for calibrating the acoustic radiation force as explained above in Sect.~\ref{subsec:setup}, while the non-zero intercept $\sigma_{0,\text{c}}=\SI{1.65}{\pascal}$ can be attributed to residual stresses trapped during the quench following the preshear \cite{mohan_2013,mohan_2015,lidon_2017a}. We checked that truncating the creep data down to about $\SI{20}{\second}$ instead of using the full time series over $\SI{300}{\second}$ does not significantly affect the fit parameters so that $\alpha$ and $\gamma^\text{(c)}_0$ are robustly defined provided the initial oscillations are fully damped.

For applied stresses below $2\sigma_{0,\text{c}}\simeq\SI{3.5}{\pascal}$, there is a competition between Andrade creep and the relaxation of residual stresses such that the Andrade law no longer accounts correctly for the data. For stresses above the yield stress $\sigma_\text{y}$, the microgel is fluidized and flows at a constant shear rate after a transient creep regime that still allows one to define an exponent $\alpha$ when the transient is not too fast \cite{divoux_2011b}.

\paragraph*{Local measurements.} Under moderate acoustic radiation forces, the sphere is first observed to jump instantaneously over a distance $r^\text{(c)}_0$ in the direction of propagation of ultrasound and then to slowly creep over time. This creep motion, strongly reminiscent of the one just described, can also be correctly accounted for by a power law, as shown in Fig.~\ref{fig:creep}(d) for $F_\text{rad}\simeq 0.5 F_\text{y}$. However, the comparison with rheological data in Fig.~\ref{fig:creep}(a) immediately shows that the relative amplitude of the sphere creep motion is much smaller than the corresponding variation in the global strain. In order to get more quantitative insights on the acoustical experiments, we fitted $r(t)$ by the Andrade behaviour 
\begin{equation}
r (t) = r^\text{(c)}_0 \left[ 1+ \left( \frac{t}{\tau} \right)^\alpha \right]\,,
\label{eq:andrade_US}
\end{equation} 
for different acoustic pressures. Using the force calibration, we can plot the Andrade exponent $\alpha$ and the instantaneous displacement $r^\text{(c)}_0$ of the sphere as a function of the acoustic radiation force $F_\text{rad}$ [see Fig.~\ref{fig:creep}(e,f)]. The exponent $\alpha$ appears to be independent of the applied force with an average value of about $0.2$, indicative of a much weaker creep motion than in macroscopic rheology. Note that in view of the experimental uncertainty, a logarithmic creep behaviour cannot be fully excluded, although we checked that better fits of the experimental data were obtained with power laws. We thus decided to account for the sphere creep motion in terms of Eq.~\eqref{eq:andrade_US} that enables a direct comparison with macroscopic rheology.

The instantaneous displacement $r^\text{(c)}_0$ is proportional to $F_\text{rad}$ at low acoustic radiation force with a negligible intercept. For forces above about $\SI{10}{\micro\newton}$ however, we observe a sublinear evolution of the instantaneous elastic deformation with $F_\text{rad}$. This striking observation is explained below in Sect.~\ref{subsec:disc_elastic_param} based on the acoustic cavity effect between the transducer and the sphere that leads to spatial variations of $F_\text{rad}$. Above the critical force $F_\text{y}=\SI{27.3}{\micro\newton}$, the sphere fluidizes the surrounding microgel and eventually moves with a non-zero velocity, as observed in Fig.~\ref{fig:example_US}(b).

\subsection{Recovery}
\label{subsec:results_recovery}

\begin{figure*}[htb]
\centering
  \includegraphics[height=10cm]{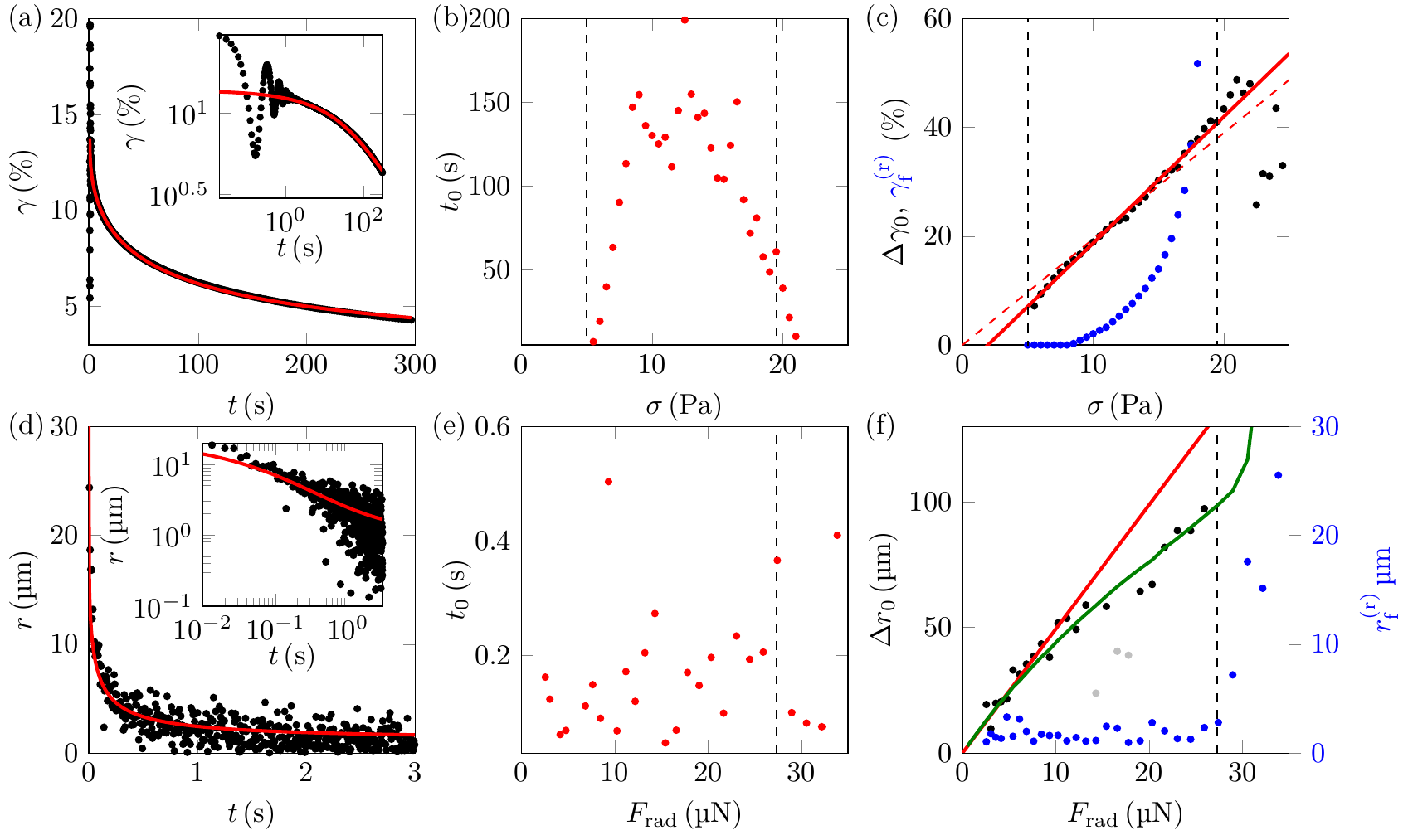}
  \caption{(a) Recovery of a 1\%~wt. Carbopol microgel as measured with a rheometer in a cone-and-plate geometry after a constant stress $\sigma = \SI{9.5}{\pascal}$ applied for $\SI{300}{\second}$ is removed at $t=0$. The red curve is a fit of the strain response $\gamma(t)$ to Eq.~\eqref{eq:powerlaw_relax_rheol} with $\gamma^\text{(r)}_0 = 13.7\%$, $\gamma^\text{(r)}_\text{f} = 1.44\%$ and $t_0=\SI{136}{\second}$ as adjustable parameters. The exponent $\alpha$ is fixed to 0.43, the value found for the corresponding creep phase shown in Fig.~\ref{fig:creep}(a). The inset shows $\gamma-\gamma^\text{(r)}_\text{f}$ as a function of time $t$ in logarithmic scales. (b)~Evolution of the characteristic time $t_0$ with the stress $\sigma$ applied prior to recovery. (c)~Instantaneous elastic recoil $\Delta \gamma_0$ (black symbols) and final deformation $\gamma_\text{f}^{\text{(r)}}$ (blue symbols) as a function $\sigma$. The red solid line is a linear fit, $\Delta \gamma_0 = (\sigma - \sigma_{0,\text{r}})/ G_\text{r}$, with $G_\text{r} = \SI{43.0}{\pascal}$ and $\sigma_{0,\text{r}}=\SI{2.0}{\pascal}$. The red dashed line corresponds to $\Delta \gamma_0=\sigma/G_\text{c}$ with $G_\text{c}=\SI{51.3}{\pascal}$. The vertical dashed lines in (b) and (c) indicate the range of stresses over which the Andrade scaling correctly accounts for the creep data, from $2\sigma_{0,\text{c}}\simeq\SI{3.5}{\pascal}$ to the yield stress $\sigma_\text{y}=19.5$~Pa.  (d)~Displacement of a polystyrene sphere embedded in a 1\%~wt. Carbopol microgel after an acoustic radiation force $F_\text{rad}=\SI{13.2}{\micro\newton}$ applied for $\SI{3}{\second}$ is removed at $t=0$. The red curve is a fit of $r(t)$ to Eq.~\eqref{eq:powerlaw_relax_US} with parameters $r^\text{(r)}_0 =\SI{31.7}{\micro\meter}$, $r^\text{(r)}_\text{f} =\SI{1.0}{\micro\meter}$ and $t_0=\SI{0.2}{\second}$. The exponent $\alpha$ is fixed to 0.18, the value found for the corresponding creep phase shown in Fig.~\ref{fig:creep}(d). The inset shows $r$ as a function of time $t$ in logarithmic scales. (e)~Evolution of the characteristic time $t_0$ with the acoustic radiation force $F_\text{rad}$ applied prior to relaxation.  (f)~Instantaneous elastic recoil $\Delta r_0$  of the sphere (black and grey symbols) and irrecoverable displacement $r_\text{f}^\text{(r)}$ (blue symbols) as a function of $F_\text{rad}$. The red line is a linear fit, $\Delta r_0 = \kappa_\text{r} F_\text{rad}$ for $F_\text{rad}<\SI{11}{\micro\newton}$, with $\kappa_\text{r} = 4.9\pm\SI{0.5}{\meter\per\newton}$. The green curve accounts for the local variations of the pressure field [see also Fig.~\ref{suppfig:recovery_local_force}(a) in Appendix~C and text in Sect.~\ref{subsec:disc_elastic_param}]. Data points that were excluded from the analysis are indicated in grey. Note the difference in the vertical scale between $\Delta r_0$ and $r_\text{f}^\text{(r)}$. Up to $F_\text{rad}=F_\text{y}$, the latter quantity remains negligible within experimental uncertainty. The vertical dashed line in (e) and (f) indicates $F_\text{y}=\SI{27.3}{\micro\newton}$.}
  \label{fig:recovery}
\end{figure*}

Figures~\ref{fig:recovery}(a) and \ref{fig:recovery}(d) display the recovery phases corresponding to Figs.~\ref{fig:creep}(a) and \ref{fig:creep}(d) respectively for macroscopic rheology and acoustical measurements. Consistently with creep measurements, we propose to describe the data in terms of power laws as a function of $\sigma$ and $F_\text{rad}$.

\paragraph*{Rheological data.} After each creep test, the microgel is allowed to relax under zero shear stress. The strain relaxation following the creep phase of Fig.~\ref{fig:creep}(a) is displayed in Fig.~\ref{fig:recovery}(a) with the origin of time $t$ taken at the beginning of the recovery phase. Inspired by recent works on power-law rheology~\cite{jaishankar_2014,aime_2018}, we choose to describe the data by the following functional form:
\begin{equation}
\gamma(t) = \gamma^\text{(r)}_\text{f} + \left(\gamma^\text{(r)}_0-\gamma^\text{(r)}_\text{f}\right) \left[ \left( 1+\frac{t}{t_0}\right)^\alpha -\left( \frac{t}{t_0}\right) ^\alpha \right]\,,
\label{eq:powerlaw_relax_rheol}
\end{equation}
where the exponent $\alpha$ is {\it fixed} to that found for the creep phase preceding relaxation. Omitting the initial inertio-elastic oscillations due to the coupling with the inertia of the geometry,  experimental results are nicely fitted by Eq.~\eqref{eq:powerlaw_relax_rheol} when relaxing from a stress lying between about $2\sigma_{0,\text{c}}$ and the yield stress $\sigma_\text{y}$ [see Fig.~\ref{fig:recovery}(a--c)].

Note that, as shown in Fig.~\ref{suppfig:recovery_exponential}(a--c) in Appendix~B, the same data can also be fitted with a sum of three exponentials given by Eq.~\eqref{eq:triple_exponential_rheol}. The quality of the resulting fits is comparable to those obtained with Eq.~\eqref{eq:powerlaw_relax_rheol}. Such an exponential relaxation could indicate that viscoelastic processes govern the recovery phase\cite{lidon_2017a}. However, the triple exponential law involves seven adjustable parameters while the power-law behaviour of Eq.~\eqref{eq:powerlaw_relax_rheol} only requires three adjustable parameters, namely the initial deformation $\gamma^\text{(r)}_0$, the final deformation $\gamma^\text{(r)}_\text{f}$ and the characteristic time $t_0$, once $\alpha$ is fixed at the value obtained in creep. Therefore, for the sake of simplicity and consistency with the power laws observed during the creep phase, we shall essentially analyze the recovery phase in terms of Eq.~\eqref{eq:powerlaw_relax_rheol} and only briefly discuss the fits by sums of exponentials.

The characteristic time $t_0$ is shown in Fig.~\ref{fig:recovery}(b) as a function of the stress $\sigma$ applied prior to recovery. Provided $\sigma$ is far enough from the upper and lower boundaries for Andrade creep behaviour, $t_0$ remains essentially constant in the range 100--150~s. Figure~\ref{fig:recovery}(c) shows the instantaneous recoil, i.e. the change of deformation immediately after the stress is set to zero $\Delta \gamma_0 = \gamma^\text{(c)}_\text{f} - \gamma^\text{(r)}_0$, where $\gamma^\text{(c)}_\text{f}$ is the strain reached at the end of the creep phase. $\Delta \gamma_0$ evolves linearly with the stress applied in the creep phase, $\Delta \gamma_0 =  (\sigma - \sigma_{0,\text{r}})/ G_\text{r}$. This shows that this initial recoil is of elastic origin and allows us to compute a second effective elastic modulus $G_\text{r} = \SI{43.0}{\pascal}$ significantly below its creep counterpart $G_\text{c} = \SI{51.3}{\pascal}$, hinting at a noticeable softening of the microgel during the creep phase.

The value of the intercept stress $\sigma_{0,\text{r}}=\SI{2.0}{\pascal}$ raises, however, some questions regarding its interpretation as the signature of residual stresses. Indeed, residual stresses should relax during the creep phase --possibly towards different final values depending on the imposed stress--, such that one would expect $\sigma_{0,\text{r}}$ to be smaller than its creep counterpart $\sigma_{0,\text{c}}=\SI{1.65}{\pascal}$. Another possible interpretation of the data lies in comparing the fitted values of $\Delta \gamma_0$ to the simple elastic prediction $\sigma/G_\text{c}$ with $G_\text{c} = \SI{51.3}{\pascal}$ [see red dashed line in Fig.~\ref{fig:recovery}(c)]. The latter prediction corresponds to a microgel where residual stresses have fully relaxed and the elastic modulus has not changed under creep. Excluding the first few points at low creep stress, this picture actually provides a good description of $\Delta \gamma_0$ up to $\sigma\simeq\SI{15}{\pascal}$. The deviation towards larger elastic recoils after the creep phase at larger stresses could then be attributed to significant damage in the microgel leading to softening.

Moreover, we observe that the final deformation $\gamma^\text{(r)}_\text{f}$ at the end of the recovery remains negligible after a creep under low stresses ($\sigma\lesssim\SI{9}{\pascal}$) and increases smoothly for higher applied stresses, reaching values as large as 50\% upon approaching the yield stress. Such irrecoverable deformation suggests that, even far below the yield stress, irreversible plastic deformation take place during the creep regime, which could induce some degree of softening. The results of exponential fits shown in Fig.~\ref{suppfig:recovery_exponential}(c) also support both interpretations, either in terms of a linear response with a smaller modulus $G_\text{r}$ and a non-zero residual stress (red solid line with $G_\text{r} = \SI{43.4}{\pascal}$ and $\sigma_{0,\text{r}}=\SI{1.3}{\pascal}$), or in terms of an elastic response with modulus $G_\text{c}$ that becomes nonlinear far below the yield stress (red dashed line). The actual response may also lie in between those two extremes. Thus, providing definite conclusions on this issue would require more investigations, e.g. by systematically varying the duration of the creep phase as in Refs.~\cite{calzolari_2017,aime_2018}. Such additional work is out of the scope of the present paper that focuses on the comparison with acoustical experiments.

\paragraph*{Local measurements.} Similarly to rheological creep and recovery experiments, the acoustic excitation is turned off for about one minute after having been applied for $\SI{3}{\second}$. During this time, the microgel relaxes and the motion of the sphere is recorded. The duration of this phase has been chosen so that the sphere reaches a stationary position before a subsequent measurement is performed at a higher acoustic intensity. The relaxation that follows the creep   phase of Fig.~\ref{fig:creep}(d) is displayed in Fig.~\ref{fig:recovery}(d).

Based on our rheological observations, we fit the sphere position $r(t)$ during recovery by the following function:
\begin{equation}
r(t) = r^\text{(r)}_\text{f} + \left(r^\text{(r)}_0-r^\text{(r)}_\text{f}\right) \left[ \left( 1+\frac{t}{t_0}\right)^\alpha -\left( \frac{t}{t_0}\right) ^\alpha \right]\,.
\label{eq:powerlaw_relax_US}
\end{equation}
Here again, $\alpha$ is fixed to the value of the exponent determined from the previous power-law fit of the sphere creep motion prior to relaxation. The relaxation occurs on timescales much shorter than in rheological measurements and experimental data rapidly involve displacements of a few micrometers that are only slightly above the noise level, which makes fitting procedures difficult. Still, Eq.~\eqref{eq:powerlaw_relax_rheol} models the experimental data correctly for $F_\text{rad}$ ranging from $\SI{3}{\micro\newton}$ to about $\SI{27}{\micro\newton}$ [see Fig.~\ref{fig:recovery}(d)]. The characteristic time $t_0$, displayed as a function the radiation force applied during the creep phase in Fig.~\ref{fig:recovery}(e), is scattered between 0.1 and 0.3~s without any obvious trend with $F_\text{rad}$. The instantaneous recoil upon release of the radiation force, $\Delta r_0 = r^\text{(c)}_\text{f} - r^\text{(r)}_0$, where $ r^\text{(c)}_\text{f}$ is the position reached by the sphere at the end of the creep phase, is displayed in Fig.~\ref{fig:recovery}(f) together with the final displacement at the end of the recovery phase $r^\text{(r)}_\text{f}$. In spite of a larger experimental scatter, $\Delta r_0$ follows an evolution similar to that of $r^\text{(c)}_0$ in Fig.~\ref{fig:creep}(f): $\Delta r_0$ is proportional to the force driving the creep phase for $F_\text{rad}\lesssim\SI{10}{\micro\newton}$, which is indicative of elastic recoil, and follows a sublinear evolution at larger forces which origin is clarified in Sect.~\ref{subsec:disc_elastic_param}. The final displacement $r^\text{(r)}_\text{f}$ remains negligible within experimental uncertainty up to $F_\text{y}=\SI{27.3}{\micro\newton}$, above which irrecoverable deformation is observed.

Last, as in the case of rheological measurements, we emphasize that the relaxation motion of the sphere when the acoustic radiation force is suppressed may also be correctly fitted by exponentials. 
An example is given in Fig.~\ref{suppfig:recovery_exponential}(d) in Appendix~B where the data of Fig.~\ref{fig:recovery}(d) are modeled by a sum of two exponentials involving the five adjustable parameters defined in Eq.~\eqref{eq:double_exponential_US}. In view of the experimental noise, fits by Eq.~\eqref{eq:powerlaw_relax_US} or by Eq.~\eqref{eq:double_exponential_US} are undistinguishable so that exponential relaxations cannot be ruled out. In the following, however, we shall favor the functional form of Eq.~\eqref{eq:powerlaw_relax_US} that involves fewer fitting parameters while being consistent with Andrade creep as discussed below in Sect.~\ref{subsec:disc_creep_rec}.

\section{Discussion}
\label{sec:discussion}

We have introduced an original technique based on the acoustic radiation force in order to explore the creep and recovery behaviour of a small sphere embedded in a microgel below its yield stress, a regime that has scarcely been investigated in the literature. Moreover, our approach allows one to investigate forces of the order of $\SI{10}{\micro\newton}$ and length scales of 10--$\SI{100}{\micro\meter}$ that are intermediate between microrheology and conventional rheology. In this Section, we present a quantitative comparison between these acoustical experiments and standard rheological characterization of the bulk microgel. We first focus on elastic properties that can be extracted from instantaneous deformation and recoil. We then discuss the similarities and differences between the power laws observed in both sets of creep and recovery experiments. Finally, we briefly address the yielding induced by the sphere motion under the effect of the acoustic radiation force.

\subsection{Instantaneous deformation and recoil in acoustical experiments}
\label{subsec:disc_elastic_param}

It is first easily checked using Eq.~\eqref{eq:force_displacement} that the linear fit of the initial sphere displacement in Fig.~\ref{fig:creep}(f), $r_0^\text{(c)} = \kappa_\text{c} F_\text{rad}$ with $\kappa_\text{c} = 6.3\pm\SI{0.3}{\meter\per\newton}$, corresponds to an elastic modulus $(6\pi a \kappa_\text{c})^{-1} = 51.7\pm\SI{2.5}{\pascal}$. Such a quantitative agreement with the modulus $G_\text{c} = \SI{51.3}{\pascal}$ inferred from macroscopic creep is fully expected since the low-intensity acoustic data were used for calibrating $F_\text{rad}$ based on this value of the elastic modulus.

A more surprising feature is the sublinear behaviour of $r^\text{(c)}_0$ as $F_\text{rad}$ increases. This behaviour can be rationalized by invoking the acoustic cavity effect that results from successive reflections of the focused ultrasonic beam incident on the sphere. Indeed, as shown in Ref.\cite{lidon_2017b}, the modulation of the local acoustic pressure amplitude is successfully modeled by a simple Fabry-P\'erot effect so that the acoustic radiation force on the sphere located at a distance $r$ from the focus of the acoustic field can be written 
\begin{equation}
F_\text{rad} (r) = \frac{\Phi(r)}{\Phi(0)} F_\text{rad} (0) \hbox{~with~} \Phi(r) = \frac{A+B \cos{[2k(r+\ell)]}}{\left(C + D \cos{[2k(r+\ell)])} \right)^2}\,,
\label{eq:cavity}
\end{equation}
where $\ell=\SI{38}{\milli\meter}$ is the focal length of the transducers and the parameters $A$, $B$, $C$ and $D$ depend only on the reflection coefficients at the sphere and transducer surfaces (see Ref.~\cite{lidon_2017b} for their expressions). The acoustic radiation force hitherto noted $F_\text{rad}$ corresponds to $F_\text{rad}(0)$ in Eq.~\eqref{eq:cavity}. The spatial dependence of $F_\text{rad}$ is shown in Fig.~\ref{suppfig:inhom_force_resolution}(a) in Appendix~D. Assuming instantaneous elastic response, one expects that the initial sphere displacement $r_0^\text{(c)}$ obeys the following local version of Eq.~\eqref{eq:force_displacement}: 
\begin{equation}
r_0^\text{(c)} = \frac{F_\text{rad}(r_0^\text{(c)})}{6\pi a G_\text{c}}\,.
\label{eq:force_displacement_cavity}
\end{equation}
We numerically solve Eqs.~\eqref{eq:cavity} and \eqref{eq:force_displacement_cavity} for various values of $F_\text{rad}(0)$ with $G_\text{c} = \SI{51.3}{\pascal}$. This leads to the green curve in Fig.~\ref{fig:creep}(f). This prediction that does not involve any fitting parameter shows excellent agreement with experimental data, thus confirming that the sublinear behaviour of $r^\text{(c)}_0$ can  be ascribed to the spatial variations of the acoustic radiation force due to the cavity effect.

As for recovery, a linear fit of the initial sphere recoil $\Delta r_0$ at low acoustic intensity leads to an effective elastic modulus $(6\pi a \kappa_\text{r})^{-1} = 66\pm\SI{7}{\pascal}$ [see red line in Fig.~\ref{fig:recovery}(f)]. Yet, at the end of the creep phase, the sphere lies at a position $r_\text{f}^\text{(c)}$ where the acoustic radiation force may differ significantly from $F_\text{rad}=F_\text{rad}(0)$ due to the above acoustic cavity effect, especially at high acoustic intensity. Therefore, for an elastic recoil, a linear scaling is expected in terms of $F_\text{rad}(r_\text{f}^\text{(c)})$ rather than $F_\text{rad}$,
\begin{equation}
\Delta r_0 = \frac{F_\text{rad}(r_\text{f}^\text{(c)})}{6\pi a G_\text{r,loc}}\,,
\label{eq:force_recoil_cavity}
\end{equation}
where $F_\text{rad}(r_\text{f}^\text{(c)})$ can be estimated directly from the experimental measurements of $r_\text{f}^\text{(c)}$ through Eq.~\eqref{eq:cavity}. Figure~\ref{suppfig:recovery_local_force}(a) in Appendix~C shows that a linear behaviour is indeed recovered for the whole range of acoustic radiation forces when $\Delta r_0$ is plotted as a function of $F_\text{rad}(r_\text{f}^\text{(c)})$. Excluding a couple of data points (shown in grey) that clearly deviate from the general trend, most probably due to fitting rather noisy data with Eq.~\eqref{eq:powerlaw_relax_US}, one gets $G_\text{r,loc}=62\pm\SI{6}{\pascal}$. The corresponding fit is plotted as a function of $F_\text{rad}$ with a green line in Fig.~\ref{fig:recovery}(f) and provides a very good description of the experimental data up to $F_\text{y}=\SI{27.3}{\micro\newton}$.

As shown in Figs.~\ref{suppfig:recovery_exponential}(f) and \ref{suppfig:recovery_local_force}(b), very similar results are obtained for the values of $\Delta r_0$ inferred from exponential fits of the recovery data. There, the effective elastic modulus is estimated to be $G_\text{r,loc}=55\pm\SI{6}{\pascal}$. This highlights the robustness of our approach to account for local variations of $F_\text{rad}$ during the creep phase. Moreover, although $\Delta r_0$ slightly depends on the choice of the fitting function and in spite of an uncertainty of about 10\%, both values of $G_\text{r,loc}$ are compatible with the elastic modulus $G_\text{c} = \SI{51.3}{\pascal}$ deduced from the creep phase. Therefore, there is no sign of a weakening of the microgel after the creep phase in our acoustical experiments, contrary to the softening and/or nonlinearity found in rheological recovery measurements. 

Finally, we note that no signature of residual stresses is observed in the acoustical experiments, contrary to the rheological characterization where non-zero intercepts $\sigma_{0,\text{c}}$ and $\sigma_{0,\text{r}}$ are reported for both the instantaneous deformation $\gamma^\text{(c)}_0$ and the instantaneous recoil $\Delta \gamma_0$. This could be explained by the fact that in the cell used for acoustical experiments, the microgel relaxation occurs at zero stress while in the rheometer, the microgel relaxes under zero deformation, which freezes residual stresses~\cite{lidon_2017b}. The more confined cone-and-plate geometry may also enhance residual stresses in rheological experiments. 

\subsection{Power-law creep and recovery}
\label{subsec:disc_creep_rec}

The most striking feature of our results on power-law creep is the significant difference between the Andrade exponent $\alpha$ found through simple shear rheology ($\alpha\simeq 0.4$) and that extracted from the sphere motion in acoustical experiments ($\alpha\simeq 0.2$). In the previous section, we have shown that the spatial variations of the acoustic radiation force $F_\text{rad}(r)$ account quantitatively for some specific features of the instantaneous displacement and recoil of the sphere. Thus, it could be argued that the spatial dependence of the driving force in acoustical experiments may explain a smaller apparent exponent $\alpha$. However, since the acoustic field is turned off after the creep phase, the acoustic cavity effect cannot play any role in recovery data besides the instantaneous recoil $\Delta r_0$. We checked that using an exponent of $\alpha\simeq 0.4$ in Eq.~\eqref{eq:powerlaw_relax_US} does not account correctly for the recovery data. This constitutes a strong argument against a smaller exponent being a mere consequence of spatial variations of $F_\text{rad}$. In Appendix~D, we further compute the sphere creep motion under a space-dependent radiation force $F_\text{rad}(r)$ and show that the spatial variations given by Eq.~\eqref{eq:cavity} are too small to modify the creep exponent for the range of positions explored by the sphere. This fully confirms that Andrade exponents in our acoustical experiments truly correspond to a much weaker creep than in macroscopic rheology.

We may suggest three interpretations of this difference in the exponent values. First, the discrepancy could come from different boundary conditions at the walls of the shear cell and at the sphere surface. Indeed, it was shown for spheres settling in a Carbopol microgel under gravity that the surface roughness affects the shape of the yielded region and the flow field around the sphere \cite{holenberg_2012}. Moreover, in some colloidal gels slippage at the walls was reported to alter --and sometimes even prevent-- the creep behaviour \cite{gibaud_2010,grenard_2014}, raising the question of bulk vs surface effects in Andrade creep. In our microgel, however, rheological measurements did not reveal any significant change in the creep behaviour when using a smooth cone and/or a bottom plate covered with sandpaper \cite{thesis_Lidon}. This makes the influence of wall slip unlikely, although the different physical chemistry of the surfaces used in rheological and in acoustical experiments may also come into play \cite{seth_2012,chan_2013}. In spite of recent experimental efforts regarding the microscopic origin of power-law creep in soft materials \cite{chan_2014,leocmach_2014,sentjabrskaja_2015,helal_2016,aime_2018b}, the question of the influence of surfaces on the creep behaviour mostly remains an open issue for soft jammed systems such as the present Carbopol microgel. 

Second, the geometries of the strain field strongly differ in the two setups. While rheological experiments involve pure shear, the strain field induced by the sphere under the effect of the acoustic radiation force is much more complex and involves both shear and extension/compression deformations. Interestingly, Andrade exponents $\alpha\simeq 0.2$ have been measured on a wide variety of living cells in a uniaxial stretching microrheometer  \cite{desprat_2005,balland_2006}. Such small values of $\alpha$ have been confirmed through active microrheology where the motion of a micronsized bead trapped in optical tweezers was followed not only under small-amplitude oscillations \cite{kollmannsberger_2011} but also through recovery experiments \cite{mandal_2016}. Let us also emphasize that indentation tests of soft materials, which essentially involve compressive stresses ahead of the indenter, most often report logarithmic creep, e.g. in calcium silicate hydrates \cite{vandamme_2009,vandamme_2013}, or at least a rather weak creep behaviour, e.g. in polyacrylamide and gelatin gels or in some biological tissues \cite{constantinides_2008,galli_2011}. Therefore, our exponents of about 0.2, which may be difficult to distinguish from logarithmic behaviour, could find their origin in the specific geometry of our acoustical ``mesorheology'' configuration that mixes shear and uniaxial extension/compression. 

Third, as reported for the very same Carbopol system in Ref.~\cite{geraud_2017}, the characteristic structural and dynamical length scales of the microgel, namely the average structure size of $\SI{2.4}{\micro\meter}$ and the ``cooperativity'' length of $\SI{11.7}{\micro\meter}$ (that corresponds to the size over which local rearrangements influence their neighborhood), are not fully negligible when compared to the radius of the sphere  $a=\SI{163}{\micro\meter}$. The weak creep behaviour observed in acoustical experiments could thus originate from the rather small sizes probed locally by the sphere within the microgel.

In summary, creep motion under the acoustic radiation force may be influenced by potential surface effects, by the complex deformation field generated by the sphere and by the extent of plastic events close to the sphere. Renewed numerical efforts are clearly needed to develop simulations of a sphere pushed through a viscoelastoplastic material below yielding and to provide decisive insights into the present observations. On the experimental side, varying the size of the sphere, changing its material and surface properties and visualizing the deformation field within the microgel could help unentangling the various effects listed above.

Finally, recovery tests also provide useful information that is often overlooked. Here, the strain and sphere relaxations are well captured by Eqs.~\eqref{eq:powerlaw_relax_rheol}  and \eqref{eq:powerlaw_relax_US} which can be derived from fractional approaches of power-law rheology, and more precisely from the fractional Maxwell (FM) model \cite{bagley_1989,schiessel_1995,jaishankar_2012,jaishankar_2014,aime_2018}. The FM model predicts that the relaxation time scale $t_0$ is given by the creep duration. This is not strictly the case in our experiments but Fig.~\ref{fig:recovery}(b) and Fig.~\ref{fig:recovery}(e) do report very different relaxation times for the strain ($t_0\sim 100$--$\SI{150}{\second}$) and for the sphere position ($t_0\sim 0.1$--$\SI{0.3}{\second}$) that indeed scale as the respective creep durations $\SI{300}{\second}$ and $\SI{3}{\second}$ of the two experiments. However, there is no reason for the FM model to hold in the case of our microgel since the viscoelastic moduli of this system do not show any clear power-law frequency dependence (see Appendix~A). Moreover, the FM model is purely linear and predicts fully recoverable strains. In contrast, in the macroscopic measurements of $\gamma_\text{f}^{\text{(r)}}$, irrecoverable deformations are reported in our microgel far below the yield stress. On the other hand, the sphere displacement always fully recovers below yielding in the acoustical setup [compare blue symbols in Fig.~\ref{fig:recovery}(c) and Fig.~\ref{fig:recovery}(f)]. This means that the response of the microgel is much more nonlinear in macroscopic measurements than in the more local acoustical experiments. Such a discrepancy is mostly due to the stronger creep applied for longer durations in the rheometer than in the acoustical setup. Yet, this observation also hints at different mechanisms of deformation in the two situations: while creep under a globally applied stress is associated with plastic irreversible events above about $\sigma_\text{y}/2$, it remains a reversible phenomenon up to yielding in the acoustical experiments. This clearly calls for more work to elucidate the basic mechanism responsible for creep in microgels.

\subsection{Local yielding induced by the sphere motion}
\label{subsec:disc_yielding}

Having discussed the features of ultrasound-induced creep below yielding, we can now analyse the value of the critical force $F_\text{y}$ at which the sphere starts to fluidize the surrounding microgel. Analytical and numerical models of viscoplastic materials by Beris {\it et al.}\cite{beris_1985} show that yielding occurs when the yield-stress parameter,
\begin{equation}
Y = \frac{2\pi a^2 \sigma_\text{y}}{F_\text{rad}}\,,
\label{eq:Beris}
\end{equation} 
\noindent falls below a critical value $Y_\text{c,th}\simeq 0.143$. However, validations of this model have always been carried out in the flowing regime which does not capture the behaviour close to the yield stress \cite{tabuteau_2007,boujlel_2012a,boujlel_2012b,putz_2008,chevalier_2013b,tokpavi_2009,hariharaputhiran_1998,sikorski_2009,jossic_2001,merkak_2006,atapattu_1995}. 

With the values of the yield stress $\sigma_\text{y}=\SI{19.5}{\pascal}$ and of the critical force $F_\text{y}=\SI{27.3}{\micro\newton}$ inferred from our experiments, Eq.~\eqref{eq:Beris} leads to a critical yield-stress parameter $Y_\text{c,exp} \simeq 0.12$. Yet, as the sphere moves away from the ultrasound focus, the applied force decreases due to the aforementioned acoustic Fabry-P{\'e}rot effect [see Fig.~\ref{suppfig:inhom_force_resolution}(a) in Appendix~D for the spatial dependence of $F_\text{rad}$]. Therefore, the estimate of $Y_\text{c,exp}$ should be corrected for the actual force exerted at the sphere location at threshold. From Fig.~\ref{fig:creep}(f), we measure that the instantaneous displacement of the sphere is $r_y \simeq \SI{120}{\micro\meter}$ for the critical force $F_\text{y}$. Based on Eq.~\eqref{eq:cavity}, this corresponds to a local force $F_\text{rad}(r_y)= \Phi(r_y)/\Phi(0)\,F_y \simeq 0.72 F_\text{y}$. Finally, using $F_\text{rad}(r_y)$ instead of $F_\text{y}$ in Eq.~\eqref{eq:Beris} yields a corrected value $Y_\text{c,exp} \simeq 0.12/0.72 \simeq 0.17$. 

Our experimental estimate of $Y_\text{c}$ is in satisfactory agreement with the theoretical viscoplastic prediction, yet about 20\% larger. It was recently shown that viscoelasticity tends to increase the critical yield-stress parameter \cite{fraggedakis_2016}. Thus, the elasticity of the microgel could explain the observed discrepancy. There again, a characterization of the deformation around the microgel at the fluidization threshold and of the subsequent flow would bring about significant progress towards a full quantitative comparison with numerical calculations in our acoustical configuration. 


\section{Conclusion and perspectives}
\label{sec:conclusion}

In a previous paper\cite{lidon_2017b}, assuming a purely elastic behaviour of Carbopol microgels in creep experiments, we estimated the acoustic radiation force exerted by a focused pressure field on a sphere embedded within the microgel. Here, we further exploited the dynamics of the sphere to get local insights on the microgel behaviour during creep and recovery experiments spanning force levels and length scales that extend the range of current microrheology techniques. The acoustical results were extensively compared with conventional rheological data.

Qualitatively, as in macroscopic rheology, the creep motion of the sphere under moderate forces is well described by a power law following instantaneous, elastic deformation, while recovery can be accounted for by a relaxation with the same exponent. However, significant  quantitative discrepancies are observed. The spatial dependence of the acoustic force explains the peculiar sublinear behaviour of the sphere instantaneous displacement and recoil with the driving force. However, it cannot account for the significantly smaller power-law creep and recovery exponents in acoustical experiments ($\alpha\simeq 0.2$) compared to standard rheology ($\alpha\simeq 0.4$). The most probable explanation for these exponents lies in the geometry of the deformation induced by the sphere within the microgel that combines shear and extension/compression at a ``mesoscale.'' These intriguing results highlight the interest of characterizing locally the creep induced by a small spherical object submitted to a constant force and of systematically comparing such a ``mesorheology'' to standard macroscopic rheology, to numerical simulations and to theoretical studies. This comparison may have important implications on the interpretation of some microrheology data and indentation tests that report similar small values of the Andrade exponent or logarithmic creep.

From a technical point of view, the present acoustical means of local rheological characterization can be improved in order to (i)~use smaller spheres with a smaller acoustic impedance mismatch relative to the surrounding material and (ii)~measure smaller displacements with a better resolution.  This should not only provide access to stiffer materials such as biopolymer gels but also strongly minimize the Fabry-P{\'e}rot cavity effect that was shown here to complicate the interpretation of the results. Moreover, the acoustic excitation can be easily modified to impose a time-dependent forcing through amplitude modulation. This should give access to local measurements of viscoelastic properties in the nonlinear regime.

From a fundamental point of view, smaller spheres will also provide information on mechanical properties at length scales comparable to the microstructure, which could turn out to be very interesting for heterogeneous materials or to investigate scales where confinement effects become predominant in conventional rheometry. The acoustic radiation force also appears as an interesting way to bridge the gap with theoretical and numerical studies. Indeed, some questions still remain open such as the relationship between the power-law creep exponent, the microstructure and the geometry of the deformation field.

\paragraph*{Acknowledgements}

The authors thank Nicolas Taberlet for technical help with the acoustical setup and Thibaut Divoux, Yo{\"e}l Forterre, Thomas Gibaud, Guillaume Ovarlez, and R{\'e}gis Wunenburger for insightful discussions. This work was funded by the Institut Universitaire de France and by the European Research Council under the European Union's Seventh Framework Programme (FP7/2007-2013)/ERC grant agreement No. 258803.
	
\section*{Appendix A. Viscoelastic measurements}

\begin{figure}[ht]
\centering
  \includegraphics[width=\columnwidth]{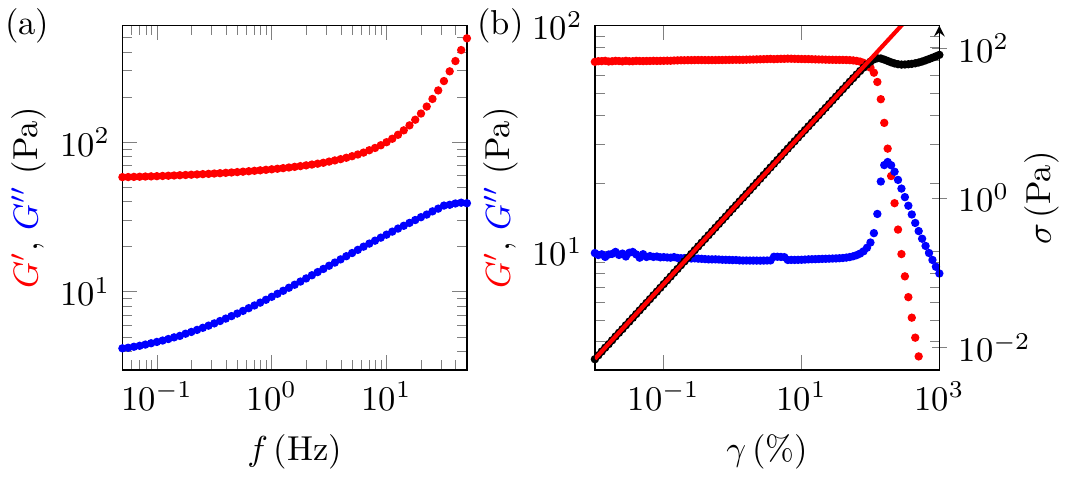}
  \caption{(a)~Elastic and viscous moduli $G'$ and $G''$ of a 1\%~wt. Carbopol microgel as a function of frequency $f$ for a sinusoidal strain of fixed amplitude $\gamma = 1 \%$. (b)~Elastic and viscous moduli (left axis) and stress amplitude (right axis) as a function of the strain oscillation amplitude $\gamma$ for a fixed frequency $f=\SI{1}{\hertz}$. The red line is a linear fit $\sigma = G_\text{osc} \gamma$ in the linear regime ($\gamma < 100\%$) with $G_\text{osc}=\SI{71}{\pascal}$.}
  \label{fig:Carbopol_viscoelasticity}
\end{figure}

Viscoelastic moduli in the linear regime are displayed in Fig.~\ref{fig:Carbopol_viscoelasticity}(a) as a function of frequency. Under small-amplitude oscillatory shear, our Carbopol microgel behaves as a viscoelastic solid with a low-frequency elastic modulus $G_0 = \SI{60}{\pascal}$ and a much smaller viscous modulus ($G'' \sim \SI{4}{\pascal}$). Figure~\ref{fig:Carbopol_viscoelasticity}(b) shows the results of a stress amplitude sweep at a fixed frequency of $\SI{1}{\hertz}$. The linear regime extends up to a strain amplitude of about $100 \%$ corresponding to a stress amplitude of $\SI{120}{\pascal}$. In the linear regime, the stress and strain amplitudes are proportional to each other and their ratio defines an elastic modulus $G_\text{osc}=\SI{71}{\pascal}$ which is fully consistent with the value of $G'(f=\SI{1}{\hertz})$ inferred from the frequency sweep of Fig.~\ref{fig:Carbopol_viscoelasticity}(a). Taking into account the fact that that these measurements were performed on a different sample loading, which may lead to variations of $\pm10\%$ of the rheological properties \cite{lidon_2017a}, these values of the elastic modulus are consistent with the ones inferred from creep and recovery in the main text.

\section*{Appendix B. Fitting the recovery phase by exponentials}

\begin{figure*}[ht]
\centering
  \includegraphics[height=10cm]{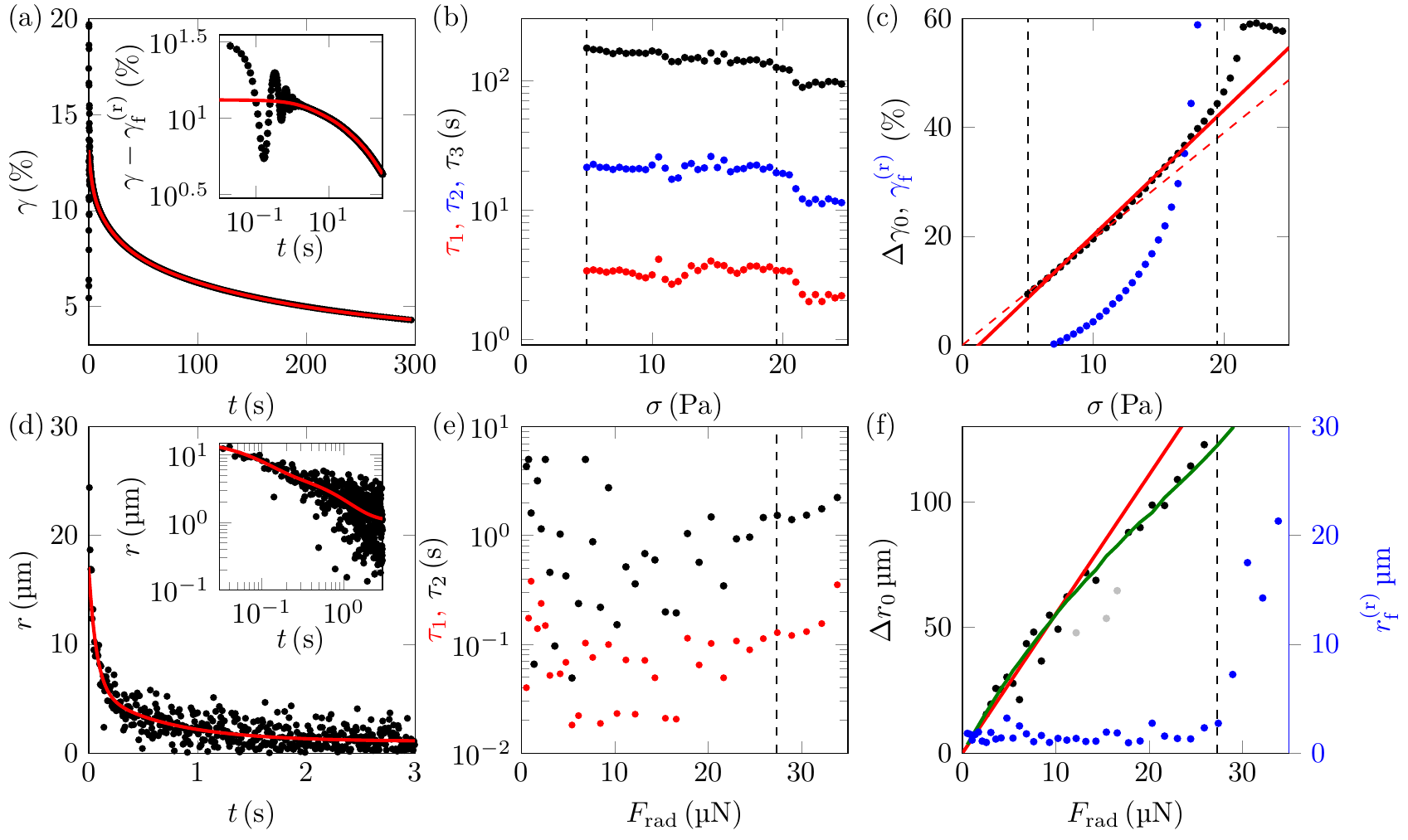}
  \caption{(a) Recovery of a 1\%~wt. Carbopol microgel as measured with a rheometer in a cone-and-plate geometry after a constant stress $\sigma = \SI{9.5}{\pascal}$ applied for $\SI{300}{\second}$ is removed at $t=0$. The red curve is a fit of the strain response $\gamma(t)$ to a triple exponential law [Eq.~\eqref{eq:triple_exponential_rheol}] with parameters  $\gamma^\text{(r)}_0 = 13.1 \%$, $A_1 = 1.9 \%$, $A_2=2.7 \%$, $A_3=5.0 \%$, $\tau_1=\SI{3.0}{\second}$, $\tau_2=\SI{20.4}{\second}$ and $\tau_3=\SI{162}{\second}$. The inset shows $\gamma-\gamma^\text{(r)}_\text{f}$ as a function of time $t$ in logarithmic scales, where $\gamma^\text{(r)}_\text{f} = \gamma^\text{(r)}_0 - A_1 - A_2 - A_3$ is the final deformation. (b)~Evolution of the three relaxation times $\tau_1$, $\tau_2$, $\tau_3$ with the stress $\sigma$ applied prior to recovery. (c)~Instantaneous elastic recoil $\Delta \gamma_0$ (black symbols) and final deformation $\gamma_\text{f}^{\text{(r)}}$ (blue symbols) as a function of $\sigma$. The red solid line is a linear fit, $\Delta \gamma_0 = (\sigma - \sigma_{0,\text{r}})/ G_\text{r}$, with $G_\text{r} = \SI{43.4}{\pascal}$ and $\sigma_{0,\text{r}}=\SI{1.3}{\pascal}$. The red dashed line corresponds to $\Delta \gamma_0=\sigma/G_\text{c}$ with $G_\text{c}=\SI{51.3}{\pascal}$. The vertical dashed lines in (b) and (c) indicate the range of stresses over which the Andrade scaling correctly accounts for the creep data, from $2\sigma_{0,\text{c}}\simeq\SI{3.5}{\pascal}$ to the yield stress $\sigma_\text{y}=19.5$~Pa. (d)~Displacement of a polystyrene sphere embedded in a 1\%~wt. Carbopol microgel after an acoustic radiation force $F_\text{rad}=\SI{13.2}{\micro\newton}$ applied for $\SI{3}{\second}$ is removed at $t=0$. The red curve is a fit of $r(t)$ to a double exponential law [Eq.~\eqref{eq:double_exponential_US}] with parameters $r^\text{(r)}_0 =\SI{17.1}{\micro\meter}$, $A_1=\SI{11.3}{\micro\meter}$, $A_2=\SI{4.78}{\micro\meter}$, $\tau_1=\SI{0.07}{\second}$ and $\tau_2=\SI{0.7}{\second}$. The inset shows $r$ as a function of time $t$ in logarithmic scales. (e)~Evolution of the two relaxation times $\tau_1$ and $\tau_2$ with the acoustic radiation force $F_\text{rad}$ applied prior to relaxation. (f)~Instantaneous elastic recoil $\Delta r_0$ of the sphere (black and grey symbols) and irrecoverable displacement $r_\text{f}^\text{(r)}= r^\text{(r)}_0 - A_1 - A_2$ (blue symbols) as a function of $F_\text{rad}$. The red line is a linear fit, $\Delta r_0 = \kappa_\text{r} F_\text{rad}$ for $F_\text{rad}<\SI{11}{\micro\newton}$, with $\kappa_\text{r} = \SI{5.9(5)}{\meter\per\newton}$. The green curve accounts for the local variations of the pressure field [see also Fig.~\ref{suppfig:recovery_local_force}(b) in Appendix~C and text in Sect.~\ref{subsec:disc_elastic_param}]. Data points that were excluded from the analysis are indicated in grey. The vertical dashed line in (e) and (f) indicates $F_\text{y}=\SI{27.3}{\micro\newton}$.}
  \label{suppfig:recovery_exponential}
\end{figure*}

\paragraph*{Rheological data.} As an alternative to Eq.~\eqref{eq:powerlaw_relax_rheol}, data from rheological recovery tests were also modeled by a triple exponential law:
\begin{equation}
\gamma(t) = \gamma^\text{(r)}_0 + A_1 (\mathrm{e}^{-t/\tau_1}-1) + A_2 (\mathrm{e}^{-t/\tau_2}-1) + A_3 (\mathrm{e}^{-t/\tau_3}-1)\,.
\label{eq:triple_exponential_rheol}
\end{equation}
Exponential laws (or sums of exponential laws) are classically used to account for viscoelastic relaxation phenomena that occur on various timescales $\tau_1$, $\tau_2$, etc.\cite{lidon_2017a} The best fit of the data following the creep test at $\sigma = \SI{9}{\pascal}$ with Eq.~\eqref{eq:triple_exponential_rheol} is shown in Fig.~\ref{suppfig:recovery_exponential}(a) for comparison with Fig.~\ref{fig:recovery}(a). The fit quality is very similar in both cases. It can be noted that the agreement is significantly better with three exponentials than with only two exponentials and that adding other exponential terms to the sum leads to poor convergence of the fitting procedure. We checked that a stretched exponential does not provide satisfying fits. 

The three characteristic times obtained with this fitting procedure are respectively of the order of $\tau_1 \simeq \SI{3}{\second}$, $\tau_2 \simeq \SI{20}{\second}$ and $\tau_3 \simeq \SI{150}{\second}$ [see Fig.~\ref{suppfig:recovery_exponential}(b)]. They do not depend significantly on the stress $\sigma$ applied during the creep phase for $2\sigma_{0,\text{c}}<\sigma<\sigma_\text{y}$. Note that the longest timescale, $\tau_3$, is of the same order as the characteristic time $t_0$ extracted from the power-law functional of Eq.~\eqref{eq:powerlaw_relax_rheol} and shown in Fig.~\ref{fig:recovery}(b).

Figure~\ref{suppfig:recovery_exponential}(c) shows the instantaneous recoil $\Delta \gamma_0$ and the final deformation at the end of the recovery $\gamma^\text{(r)}_\text{f} = \gamma^\text{(r)}_0 - A_1 - A_2 - A_3$ as a function of $\sigma$. These quantities display dependencies on $\sigma$ that are very similar to those in Fig.~\ref{fig:recovery}(c). In particular, $\Delta \gamma_0$ evolves linearly with $\sigma$ and the corresponding effective elastic modulus, $G_\text{r} =\SI{43.4}{\pascal}$, is very close to that found from fits using Eq.~\eqref{eq:powerlaw_relax_rheol}. The final deformation $\gamma^\text{(r)}_\text{f}$, however, seems to increase smoothly with $\sigma$ without showing any fully reversible regime where $\gamma^\text{(r)}_\text{f}=0$. These somewhat unphysical values of  $\gamma^\text{(r)}_\text{f}$ at low stresses differ from the results based on Eq.~\eqref{eq:powerlaw_relax_rheol} that clearly show fully recoverable deformations for $\sigma\lesssim \SI{9}{\pascal}$. This could mean that fitting by exponentials is less adequate than the power-law approach at least to model recovery after a small stress has been applied. 

More quantitatively, $\chi^2$ tests give very similar values for the power-law approach used in Sect.~\ref{subsec:results_recovery} and for the present exponential model, which does not allow us to discriminate between them. As uncertainty analysis is not straightforward in our experiments, we could not reliably perform more sophisticated statistical tests of the fit quality, such as computing the reduced $\chi^2$ statistics in order to advocate a larger number of fitting parameters.

\paragraph*{Local measurements.} The sphere position $r(t)$ was also fitted by a sum of exponentials. However, due to a larger experimental noise level, fitting procedures with a large number of adjustable parameters as in Eq.~\eqref{eq:triple_exponential_rheol} do not converge. The most reliable results were obtained by using a sum of only two exponential relaxations:
\begin{equation}
r(t) = r^\text{(r)}_0 + A_1 (\mathrm{e}^{-t/\tau_1}-1) + A_2 (\mathrm{e}^{-t/\tau_2}-1)\,.
\label{eq:double_exponential_US}
\end{equation}
 The two relaxation times $\tau_1$ and $\tau_2$ are shown in Fig.~\ref{suppfig:recovery_exponential}(e). They are rather scattered at small values of the acoustic radiation force but for $F_\text{rad}\gtrsim\SI{15}{\micro\newton}$, they clearly define two distinct values of the order of $\SI{0.1}{\second}$ and $\SI{1}{\second}$ that do not show any significant trend with $F_\text{rad}$. Here again, the longest timescale $\tau_2$ is comparable to the characteristic time $t_0$ shown in Fig.~\ref{fig:recovery}(e).

Figure~\ref{suppfig:recovery_exponential}(f) shows the instantaneous recoil upon release of the radiation force, $\Delta r_0$, and the final displacement $r^\text{(r)}_\text{f} = r^\text{(r)}_0 - A_1 - A_2$. Up to the uncertainty due to the fitting procedure, these parameters behave similarly to those extracted from Eq.~\eqref{eq:powerlaw_relax_US} and shown in Fig.~\ref{fig:recovery}(f). Moreover, as for rheological data, the $\chi^2$ test does not allow us to favour this exponential model over the power-law model used in the main text.

\section*{Appendix C. Accounting for local variations of the pressure amplitude in recovery experiments}

\begin{figure}[ht]
\centering
  \includegraphics[width=\columnwidth]{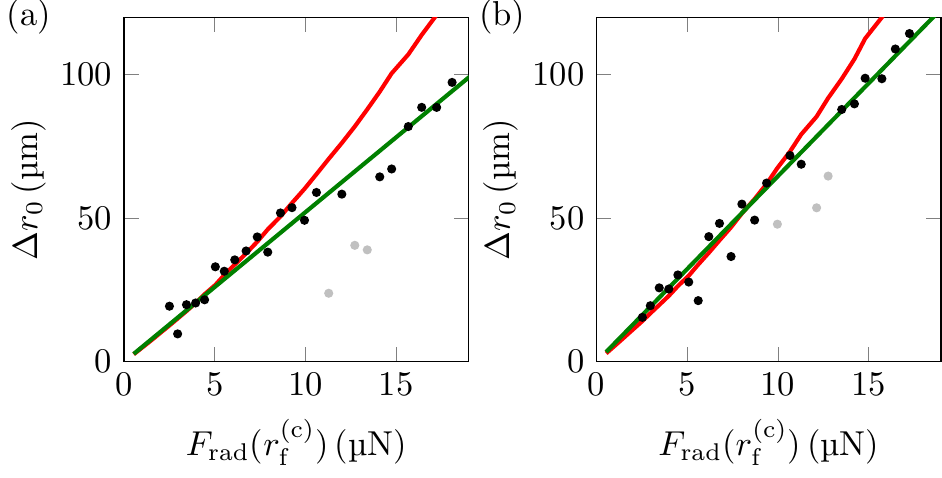}
  \caption{Instantaneous sphere recoil $\Delta r_0$ plotted as a function of the local acoustic radiation force $F_\text{rad}(r_\text{f}^\text{(c)})$ at the position $r_\text{f}^\text{(c)}$ reached by the sphere at the end of the creep phase. (a) Data extracted from fits to the power-law relaxation of Eq.~\eqref{eq:powerlaw_relax_US}. (b) Data extracted from fits to the double exponential relaxation of Eq.~\eqref{eq:double_exponential_US}. The green lines in (a) and (b) are the best linear fits to Eq.~\eqref{eq:force_recoil_cavity}, with $G_\text{r,loc}=62\pm\SI{6}{\pascal}$ and $G_\text{r,loc}=55\pm\SI{6}{\pascal}$ respectively, when the data points indicated in grey are excluded from the fits. The red curves in (a) and (b) correspond to the linear fits of $\Delta r_0$ {\it vs} $F_\text{rad}$ also shown in red in Fig.~\ref{fig:recovery}(f) and in Fig.~\ref{suppfig:recovery_exponential}(f) respectively.}
  \label{suppfig:recovery_local_force}
\end{figure}

Figure~\ref{suppfig:recovery_local_force} shows that a linear behaviour is recovered up to the fluidization threshold when the instantaneous sphere recoil $\Delta r_0$ is plotted as a function of $F_\text{rad}(r_\text{f}^\text{(c)})$ rather than $F_\text{rad}$. This is observed both for fits to the power-law relaxation function of Eq.~\eqref{eq:powerlaw_relax_US} and for fits to the double exponential relaxation of Eq.~\eqref{eq:double_exponential_US}. This shows that the sublinear trends observed in Fig.~\ref{fig:recovery}(f) and Fig.~\ref{suppfig:recovery_exponential}(f) can be unambiguously attributed to the effect of the spatial dependence of $F_\text{rad}$ on the purely elastic recoil.

\section*{Appendix D. Testing for the influence of spatial variations of the pressure amplitude on the Andrade exponent}

\begin{figure}[ht]
\centering
  \includegraphics[width=\columnwidth]{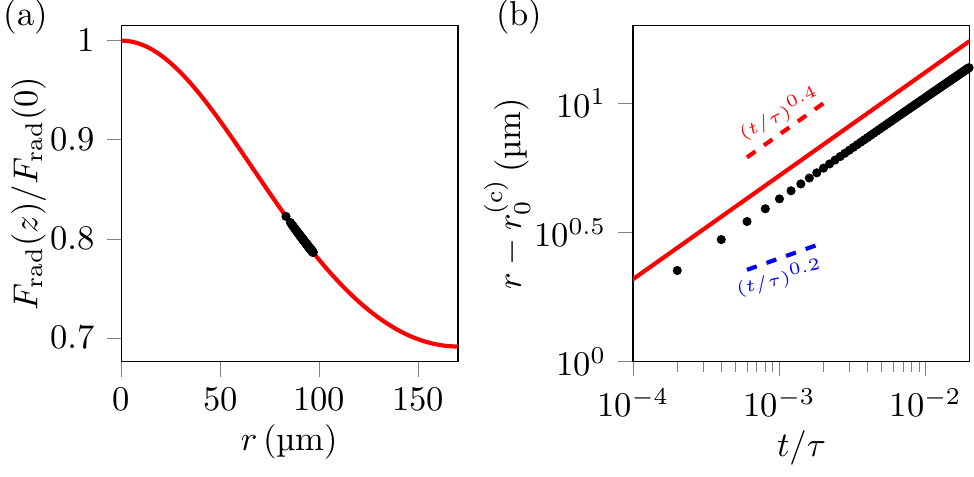}
  \caption{(a)~Local value of the acoustic radiation force $F_\text{rad}(r)$ at a distance $r$ from the focus of the acoustic field and normalized by the value at focus $F_\text{rad}(0)$. The analytic expression for the spatial dependence of $F_\text{rad}(r)$ is given in Eq.~\eqref{eq:cavity}. The black dots show the range of positions and forces that correspond to the sphere creep motion for $F_\text{rad}=\SI{13.2}{\micro\newton}$ as shown in Fig.~\ref{fig:creep}(d). (b)~Creep motion $r-r_0^\text{(c)}$ under the space-dependent force shown in (a) assuming a kernel with a power-law exponent $\alpha=0.4$. Black dots show the solution of Eq.~\eqref{eq:position_force_variable_3} as a function of $t/\tau$. The parameters of the calculation have been tuned so that the amplitude of the creep motion reaches that observed experimentally for $F_\text{rad}=\SI{13.2}{\micro\newton}$, i.e. about $\SI{14}{\micro\meter}$. The red curve is the solution $r_\text{hom}(t)$ in the case of a homogeneous force. The red and blue dotted lines correspond to power laws of exponents 0.4 and 0.2 respectively.}	
  \label{suppfig:inhom_force_resolution}
\end{figure}

According to linear response theory, the sphere position is given by
\begin{equation}
    r(t)=\int_{-\infty}^t \dot{M}(t-t')F(t')\text{d}t'\,,
\label{eq:position_memory_function}
\end{equation}
where $F(t)$ is the force applied on the sphere at time $t$. For the present power-law creep, the convolution kernel is $M(t)=\kappa[1+(t/\tau)^\alpha]H(t)$ with $H(t)$ the Heaviside function and $\kappa$ the proportionality factor such that the instantaneous elastic jump under a constant force $F_\text{rad}$ is $r_0^\text{(c)}=\kappa F_\text{rad}$. In the following, we shall assume that the creep exponent of the sphere under a {\it spatially homogeneous} force is the same as that found in rheological measurements, {\it i.e.} we take $\alpha=0.4$ as the exponent for $M(t)$. In other words, if the acoustic radiation force did not depend on the sphere position, we assume that the sphere response would be given by $r_\text{hom}(t) =r_0^\text{(c)} [ 1+ (t/\tau)^\alpha ]$ with $\alpha=0.4$.

For a {\it space-dependent} force applied at $t=0$, i.e. $F(t)=F_\text{rad}[r(t)]H(t)$, Eq.~\eqref{eq:position_memory_function} reads 
\begin{equation}
    r(t)=\int_0^t \dot{M}(t-t')F_\text{rad}[r(t')]\text{d}t'\,.
\label{eq:position_force_variable_2}
\end{equation}
Solving directly such a nonlinear Volterra integral equation requires involved numerical methods which are beyond the scope of the present paper \cite{brunner_1986,cardone_2018}. Yet, noting that the range of positions explored by the sphere during creep is very limited [see Fig.~\ref{suppfig:inhom_force_resolution}(a)], we use the following iterative method. Rewriting Eq.~\eqref{eq:position_force_variable_2} as
\begin{equation}
    r(t) = r_\text{hom}(t) + \kappa \int_0^t \frac{\alpha(t-t')^{\alpha-1}}{\tau^\alpha} \left[ F_\text{rad}[r(t')] - F_\text{rad} \right] \text{d}t'\,,
\label{eq:position_force_variable_3}
\end{equation}
we start with $r_0(t)=r_\text{hom}(t)$ and introducing the reduced variable $u=t/\tau$, we iteratively compute $r_{i+1}(u)$ from $r_i(u)$ as
\begin{equation}
    r_{i+1}(u) = r_\text{hom}(u) + \kappa \int_0^u \alpha(u-u')^{\alpha-1} \left[ F_\text{rad}[r_i(u')] - F_\text{rad} \right] \text{d}u'\,.
\label{eq:position_force_variable_4}
\end{equation}

In practice, this process converges very fast and the solution found after a couple of iterations in the case of $F_\text{rad}=\SI{13.2}{\micro\newton}$ is shown in Fig.~\ref{suppfig:inhom_force_resolution}(b). The exponent of $r(t)$ is indistinguishable from 0.4. Actually, the solution only slightly differs from the case of a spatially homogeneous force, $r_\text{hom}(t)$. Therefore, the spatial variations of the acoustic radiation force cannot account for a decrease in the creep exponent from 0.4 in the macroscopic rheology to 0.2 in the sphere motion. 

Similar results are obtained when starting from a kernel with exponent 0.2: $\alpha$ remains unaffected by the variations of $F_\text{rad}(r)$. Also note that the effect of $F_\text{rad}(r)$ is not necessarily larger for higher acoustic radiation forces as the sinusoidal shape of $F_\text{rad}(r)$ mitigates the influence of larger creep motions [see Fig.~\ref{suppfig:inhom_force_resolution}(a)]. We conclude that during its creep motion, the sphere spans a range of positions that is so small that the effect of the spatial variations of the driving force on the Andrade exponent can be completely neglected up to the fluidization threshold. These spatial variations only have a significant influence on the instantaneous jump and recoil of the sphere at the very beginning of the creep and recovery phases.


\end{document}